%% file: spacehbt.tex
\documentclass[superscriptaddress,preprint,secnumarabic,amssymb,nobibnotes,aps,prc]{revtex4}
\usepackage{graphicx}
\usepackage{bm}
\input{define}

\begin{document}
\title{Space-time evolution and HBT analysis
of relativistic heavy ion collisions 
in a chiral $SU(3) \times SU(3)$ model.}
\author{D. Zschiesche}
\affiliation{Institut f\"ur Theoretische Physik,
        Postfach 11 19 32, D-60054 Frankfurt am Main, Germany}

\author{S. Schramm}
\affiliation{Argonne National
Laboratory, 9700 S. Cass Avenue, Argonne IL 60439, USA}

\author{H.~St\"ocker}
\affiliation{Institut f\"ur Theoretische Physik,
        Postfach 11 19 32, D-60054 Frankfurt am Main, Germany}

\author{W.~Greiner}
\affiliation{Institut f\"ur Theoretische Physik,
        Postfach 11 19 32, D-60054 Frankfurt am Main, Germany}

\begin{abstract}
The space-time dynamics and pion-HBT radii in central heavy
ion-collisions at CERN-SPS and BNL-RHIC are investigated 
within a hydrodynamic simulation.
The dependence of the dynamics and the HBT-parameters on the EoS
is studied with different parametrizations of a 
chiral SU(3) $ \sigma-\omega$ model. 
The selfconsistent collective expansion
includes the effects of effective hadron masses, generated
by the nonstrange and strange scalar condensates. Different chiral EoS
show different types of phase transitions and even a crossover. 
The influence of the
order of the phase transition and of the latent heat 
on the space-time dynamics and pion-HBT radii is studied. 
A small latent heat, i.e.\ a weak
first-order chiral phase transition, or a smooth crossover lead
to distinctly different
HBT predictions than 
a strong first order phase transition.
A quantitative description of the data, both at SPS energies
as well as at RHIC energies, appears difficult to achieve within 
the ideal hydrodynamic approach using the SU(3) chiral EoS.
A strong first-order quasi-adiabatic chiral phase transition 
seems to be disfavored by the pion-HBT data from CERN-SPS and
BNL-RHIC.
\end{abstract}
\pacs{}
\maketitle
\section{Introduction}

General theoretical arguments~\cite{Pisarski:1984ms} and lattice QCD
simulations~\cite{lattice_chi}
predict the occurrence of a transition of strongly interacting matter
to a state where chiral symmetry is (approximately) restored.
Since Bose-Einstein correlations in multiparticle production
processes~\cite{Goldhaber} provide valuable information on the 
space-time dynamics of fundamental interactions~\cite{Shuryak:1973kq}, 
correlations of identical pions
produced in high energy collisions of heavy ions may provide
information on the characteristics of that phase
transition (for a review on QGP signatures, see \cite{Bass:1999vz}).
For recent reviews on this topic we refer
to~\cite{reviews,wiedemannrep}. 

In particular, a first order phase transition
leads to a prolonged hadronization time as compared to a cross-over
or a hadron gas with no symmetry restoration, and has been related to unusually
large Hanbury-Brown--Twiss (HBT) radii~\cite{pratt86,schlei,RiGy}.
The coexistence of hadrons and QGP reduces the ``explosivity'' of the
high-density matter before hadronization, prolonging the emission duration of 
pions~\cite{pratt86,schlei,RiGy}.
This phenomenon should then depend on the
critical temperature $T_c$ and the latent heat of the
transition. Typically, calculations assuming a first-order phase
transition are carried out with an equation of state (EoS) derived from
matching the bag model with an ideal hadron gas model, 
 for which the latent heat of the transition is
large~\cite{schlei,RiGy}.
Consequently, the predicted HBT radii were large.

 Here, we 
consider also the case of a more weakly first-order transition with
small latent heat and study the influence on 
the space time characteristics of the expansion and on the HBT
radii. Furthermore, we 
perform explicit calculations for a
smooth transition (crossover) at high temperatures, 
and discuss the resulting
pion HBT radii. Such a scenario was considered in \cite{RiGy}, however
without explicit reference to chiral symmetry restoration and dynamical
hadron masses.

To investigate the space-time dynamics
and the influences of different types of phase transitions,  
hydrodynamic expansion with an EoS obtained from a chiral $SU(3)_L \times
SU(3)_R \,\sigma-\omega$ model is considered. The equations of fluid
dynamics describe the
collective evolution of the system, while the chiral $SU(3) \times
SU(3)$ model yields the underlying equation of state. 
Thus, as the hot and dense central region expands both in the longitudinal
and transverse directions, the hadrons approach their vacuum masses.
The initial excitation energy is converted into both, collective
flow \emph {and} massive hadrons. This purely hadronic model
describes successfully nuclear matter ground state properties, finite
nuclei and hadron masses in the vacuum \cite{paper3,springer}. 
Furthermore, it exhibits different kinds of 
high temperature transitions, depending on the set of parameters.
Using the various
equations of state in a hydrodynamic simulation should discriminate between 
the different phase transition scenarios. Since the model only
contains hadronic degrees of freedom, we only test the influence of
the chiral phase transition but not of the deconfinement phase
transition. In any case,
the main effect as far as collective expansion is concerned,
is due to the difference in the latent heat for the transition, irrespective
of its microscopic origin.

This article is organized as follows. In section~\ref{the_model} we
present our model. In particular, in~\ref{sec_hydro} we discuss
ideal relativistic hydrodynamics, and in~\ref{sec_EOS} we refer to our
equations of state. Section~\ref{sec_results} shows our main results for
the space-time evolution and the pion HBT radii. We summarize and conclude
in section~\ref{sec_summary}. Throughout the manuscript, we employ
natural units $c=\hbar=k_B=1$.

\section{Model Description} \label{the_model}

\subsection{Scaling Hydrodynamics} \label{sec_hydro}

Ideal 
Hydrodynamics is defined by (local) energy-momentum and net charge
conservation~\cite{Landau},
\be \label{Hydro}
\partial_\mu T^{\mu\nu}=0 \quad,\quad
\partial_\mu N_i^{\mu}=0\quad.
\ee
$T^{\mu\nu}$ denotes the energy-momentum tensor, and $N_i^{\mu}$ the
four-current of the $i$th conserved charge. We will explicitly consider only
one such conserved charge, the net baryon number. We implicitly assume
that the local densities of all other charges which are conserved on
strong-interaction time scales,
e.g.\ strangeness, charm, and electric charge, vanish. The
corresponding four-currents are therefore identically zero, cf.\
eq.~(\ref{idfluid}), and the conservation equations are trivial.

For ideal fluids, the energy-momentum tensor and the net baryon current
assume the simple form~\cite{Landau}
\be \label{idfluid}
T^{\mu\nu}=\left(\epsilon+p\right) u^\mu u^\nu -p g^{\mu\nu}
\quad,\quad
N_B^{\mu}=\rho_B u^\mu \quad,
\ee
where $\epsilon$, $p$, $\rho_B$ are energy density, pressure, and net baryon
density in the local rest frame of the fluid, which is defined by
$N_B^\mu=(\rho_B,\vec{0})$. $g^{\mu\nu}={\rm diag}(+,-,-,-)$ is the metric
tensor, and $u^\mu=\gamma(1,\vec{v})$ the four-velocity of the fluid
($\vec{v}$ is the three-velocity and $\gamma=(1-\vec{v}^2)^{-1/2}$ the
Lorentz factor). The system of partial differential
equations~(\ref{Hydro}) is closed by choosing an equation of state (EoS)
in the form $p=p(\epsilon,\rho_B)$, cf.\ below.

For simplicity, we assume cylindrically symmetric transverse expansion with a
longitudinal scaling flow profile, $v_z=z/t$~\cite{Bj}.
This should be a reasonable first approximation
for central collisions at high energy (such as at CERN-SPS and BNL-RHIC
energies), and around midrapidity. 
A quantitative comparison to experimental data, which we postpone to
a future publication, should however analyze the effects due to coupling of
longitudinal and transverse flows around midrapidity. At least up to
CERN-SPS energies, $\sqrt{s}\sim20A$~GeV, such a coupling was shown to
exist~\cite{longtranscouple}.

The hydrodynamic equations of motion are solved on a discretized
space-time grid ($\Delta r_T=R_T/100=0.06$~fm, $\Delta \tau=0.99\Delta r_T$)
by employing the RHLLE algorithm as described and tested in~\cite{RiGy,RiBe}.
We have checked that the algorithm accurately conserves total
energy and baryon number, and that profiles of rarefaction
and shock waves are reproduced accurately for various initial
conditions~\cite{DumRi,RiBe,RiGy2}.

As already mentioned above, eqs.~(\ref{idfluid}), we assume a perfect,
i.e. non-dissipative, relativistic fluid. In principle, it is possible to
calculate the transport coefficients from the Lagrangean of our
model~\cite{paper3,springer}. (For example, various transport coefficients
have been computed in the symmetry broken phase based on the assumption of
an ideal gas of hadrons~\cite{Prakash:1993bt}.)
Also, dynamical simulations indicate that dissipation strongly affects
the pion correlation functions at small relative momentum, and thus the
deduced HBT radii~\cite{soff}. Quantitative comparisons to experimental
data should therefore account for dissipative effects. On the other hand,
the purpose of this paper is to explore the effects from varying the
latent heat and the order of the phase transition. In that vein, we
can leave aside the great technical and principal difficulties 
related to a treatment
of dissipation in dynamical simulations~\cite{muronga},
and give an impression of the largest
possible effects of varying the phase transition parameters that can be
expected. This will also allow for a comparison to previous results for
the pion HBT correlation functions, which employed ideal fluid dynamics
with an EoS derived from the bag model \cite{schlei,RiGy}.

\subsection{Equations of state from a chiral $SU(3) \times SU(3)$
model} \label{sec_EOS}
To close the system of coupled equations of hydrodynamics, an equation of
state (EoS) has to be specified. Lattice QCD predicts
chiral symmetry restoration at 
a critical temperature of $T_c=140-170$~MeV~\cite{lattice_chi,Laerm}
(for $\rho_B=0$).
We obtain the equation of state from a chiral   
$SU(3) \times SU(3) \sigma-\omega$ model
that was discussed in detail in 
\cite{paper3,springer}. We will briefly introduce the
model here:
consider a relativistic field theoretical model of 
baryons and mesons based on a nonlinear realization of 
chiral symmetry and broken scale invariance. The general form of the
Lagrangean is:
\be
\label{lagrange}
{\cal L} = {\cal L}_{\mathrm{kin}}+\sum_{M=X,Y,V,{\cal A},u}{\cal
L}_{\mathrm{BM}}
+{\cal L}_{\mathrm{vec}}+{\cal L}_{\mathrm{VP}}
-{\cal V}_{0}-{\cal V}_{\mathrm{SB}} .\no
\ee
${\cal L}_{\mathrm{kin}}$ is 
the kinetic energy term, ${\cal L}_{\mathrm{BM}}$ includes the  
interaction terms of the different baryons with the various spin-0 and spin-1 
mesons. The baryon masses are generated by both, the nonstrange
$\sigma$ (${<q\bar{q}>}$)  
 and the strange $\zeta$ (${<s\bar{s}>}$) scalar condensate. 
X,Y,V,A,u stand for scalar octet, scalar singlet, vector, axial vector
and pseudoscalar mesons respectively.
${\cal L}_{\rm{VP}}$ contains the interaction terms 
of vector mesons with pseudoscalar mesons. 
${\cal L}_{\rm{vec}}$ generates the masses of the spin-1 mesons through 
interactions with spin-0 mesons, and ${\cal V}_{0}$ gives the meson-meson 
interaction terms which induce the spontaneous breaking of chiral symmetry.
It also includes a scale-invariance breaking logarithmic potential. Finally, 
${\cal V}_{\mathrm{SB}}$ introduces an explicit symmetry breaking of the
U(1)$_A$, the SU(3)$_V$, and the chiral symmetry. 
All these terms have been discussed in detail in \cite{paper3,springer}.\\
\label{mfa}
The hadronic matter
properties at finite density and temperature are studied in 
the mean-field approximation, i.e. the meson field operators are
replaced by their expectation values and the fermions 
are treated as quantum mechanical one-particle operators \cite{serot97}. 
After performing these approximations, the Lagrangean (\ref{lagrange}) 
becomes
\begin{eqnarray*}
{\cal L}_{BM} &=& -\sum_{i} \overline{\psi_{i}}[g_{i 
\omega}\gamma_0 \omega^0 
+g_{i \phi}\gamma_0 \phi^0 +m_i^{\ast} ]\psi_{i} \\ \no
{\cal L}_{vec} &=& \frac{ 1 }{ 2 } m_{\omega}^{2}\frac{\chi^2}{\chi_0^2}\omega^
2  
 + \frac{ 1 }{ 2 }  m_{\phi}^{2}\frac{\chi^2}{\chi_0^2} \phi^2
+ g_4^4 (\omega^4 + 2 \phi^4)\\
{\cal V}_0 &=& \frac{ 1 }{ 2 } k_0 \chi^2 (\sigma^2+\zeta^2) 
- k_1 (\sigma^2+\zeta^2)^2 
     - k_2 ( \frac{ \sigma^4}{ 2 } + \zeta^4) 
     - k_3 \chi \sigma^2 \zeta \\ 
&+& k_4 \chi^4 + \frac{1}{4}\chi^4 \ln \frac{ \chi^4 }{ \chi_0^4}
 -\frac{\delta}{3} \chi^4 \ln \frac{\sigma^2\zeta}{\sigma_0^2 \zeta_0} \\ \no
{\cal V}_{SB} &=& \left(\frac{\chi}{\chi_0}\right)^{2}\left[m_{\pi}^2 f_{\pi} 
\sigma 
+ (\sqrt{2}m_K^2 f_K - \frac{ 1 }{ \sqrt{2} } m_{\pi}^2 f_{\pi})\zeta 
\right] , 
\end{eqnarray*}
with $m_i$ the effective mass of the baryon $i$  
($i=N,\Lambda,\Sigma,\Xi,\Delta,\Sigma^\ast,\Xi^\ast,\Omega$). 
$\sigma$
and $\zeta$ correspond to  
the scalar condensates, $\omega$ and $\phi$ represent
the iso-singlet non-strange and the strange vector field,
respectively, 
and $\chi$ is
the dilaton field, which can be viewed as representing the
 effects of the gluon condensate. 
In this work we will use the frozen glueball approximation, i.e.
adopt the dilaton field as constant. In the current form of the model
this makes sense, since the glueball field does not change strongly with
temperature and density. In a forthcoming work we will investigate the
consequences of a stronger coupling of the glueball field to the
scalar fields.

The thermodynamical potential of the grand canonical 
ensemble $\Omega$ per volume $V$ 
at a given chemical potential $\mu$ and temperature $T$ can 
be written as:
\begin{eqnarray*}
   \frac{\Omega}{V}&=& -{\cal L}_{vec} + {\cal V}_0 + {\cal V}_{SB}
-{\cal V}_{vac} \\
&-& \frac 1T \sum_i \frac{\gamma_i }{(2 \pi)^3}
\int d^3k \left[\ln{\left(1+e^{-\frac 1T[E^{\ast}_i(k)-\mu^{\ast}_i]}\right)}
\right] \\
&+& \frac 1T \sum_j \frac{\gamma_j }{(2 \pi)^3}
\int d^3k \left[\ln{\left(1-e^{-\frac 1T[E^{\ast}_j(k)-\mu_j]}\right)}
\right] 
\end{eqnarray*}
The vacuum energy ${\cal V}_{vac}$ (the potential at $\rho=0$) 
has been subtracted in 
order to get a vanishing total vacuum energy. $\gamma_i$ denote the
fermionic and $\gamma_j$ the bosonic  
spin-isospin degeneracy factors.
The single particle energies are 
$E^{\ast}_i (k) = \sqrt{ k_i^2+{m_i^*}^2}$, with 
$m_i^* = m_i^*(\sigma,\zeta)$ 
(see \cite{paper3,springer}). The effective baryonic chemical
potentials read
 $\mu^{\ast}_i = \mu_i-g_{i\omega} \omega-g_{i\phi} \phi$ 
with $\mu_i = (n_q^i - n_{\bar q}^i) \mu_q +(n_s^i - n_{\bar s}^i) \mu_s$
and the
mesonic chemical potentials read 
$\mu_j = (n_q^j - n_{\bar q}^j) \mu_q +(n_s^j - n_{\bar s}^j) \mu_s$.
$n_q^i$, $n_{\bar q}^i$,$n_s^i$ and $n_{\bar s}^i$ 
denote the number of consituent $q$, $\bar q$, $s$ and $\bar{s}$ quarks in
particle species $i$, respectively. 
\\ 
The mesonic fields are determined by extremizing $\frac{\Omega}{V}(\mu,T)$:
The density of particle $i$ can be calculated by differentiating
$\Omega$ with respect to the corresponding chemical potential $\mu_i$.
This yields:
\begin{eqnarray*}
\rho_i = \gamma_i
\int_0^\infty \frac{d^3 k}{(2 \pi)^3}
\left[
\frac 1{\exp{[(E_i^\ast-\mu_i^\ast)/T]\pm1}}
\right] 
\end{eqnarray*}
The net density of particle species $i$ is given by 
$\rho_i - \bar{\rho_i}$.
The energy density and the pressure  follow from the Gibbs--Duhem relation,
$\epsilon = \Omega/V+ T S + \mu_i \rho^i$ and $p= - \Omega/V$.
%
%
%
The model shows a phase transition or a crossover 
around $T_c = 150 \rm{MeV}$.
Since there are only hadronic degrees of freedom in the model, this phase
transition is of purely hadronic nature, i.e. the strong increase of
the scalar density reduces the masses of the baryons, which in turn
again increases the scalar density (compare e.g. to \cite{Theis}).
  
The characteristics (e.g. the order, the latent heat) of the various  
phase transitions depend on the chosen parameters
and on the considered degrees of freedom.
We will use three different 
parameter sets, which differ in their  
treatment of the baryon resonances. 
This  
leads to different predictions concerning the behavior of
hot hadronic matter. In parameter set CI the baryon 
decuplet is neglected, and the only degrees of freedom in the system are
the members of the (anti)-baryon octet, 
the pseudoscalar meson nonet and 
the vector meson nonet. 
In parameter set CII and CIII we include
the (anti)-baryon decuplet. This increases the number of degrees of freedom by
$80$. The  
parameter sets CII and CIII differ in the 
treatment of the strange spin-$\frac 32$
resonances. In parameter set CII an additional explicit symmetry 
breaking for the baryon resonances along the hypercharge direction,  
as desribed in \cite{paper3} for the baryon octet, is included. This is 
neglected in parameter set CIII. 

\begin{figure}[h]
\vspace*{-1cm}
\centerline{\parbox[b]{8cm}{
\includegraphics[width = 9cm]{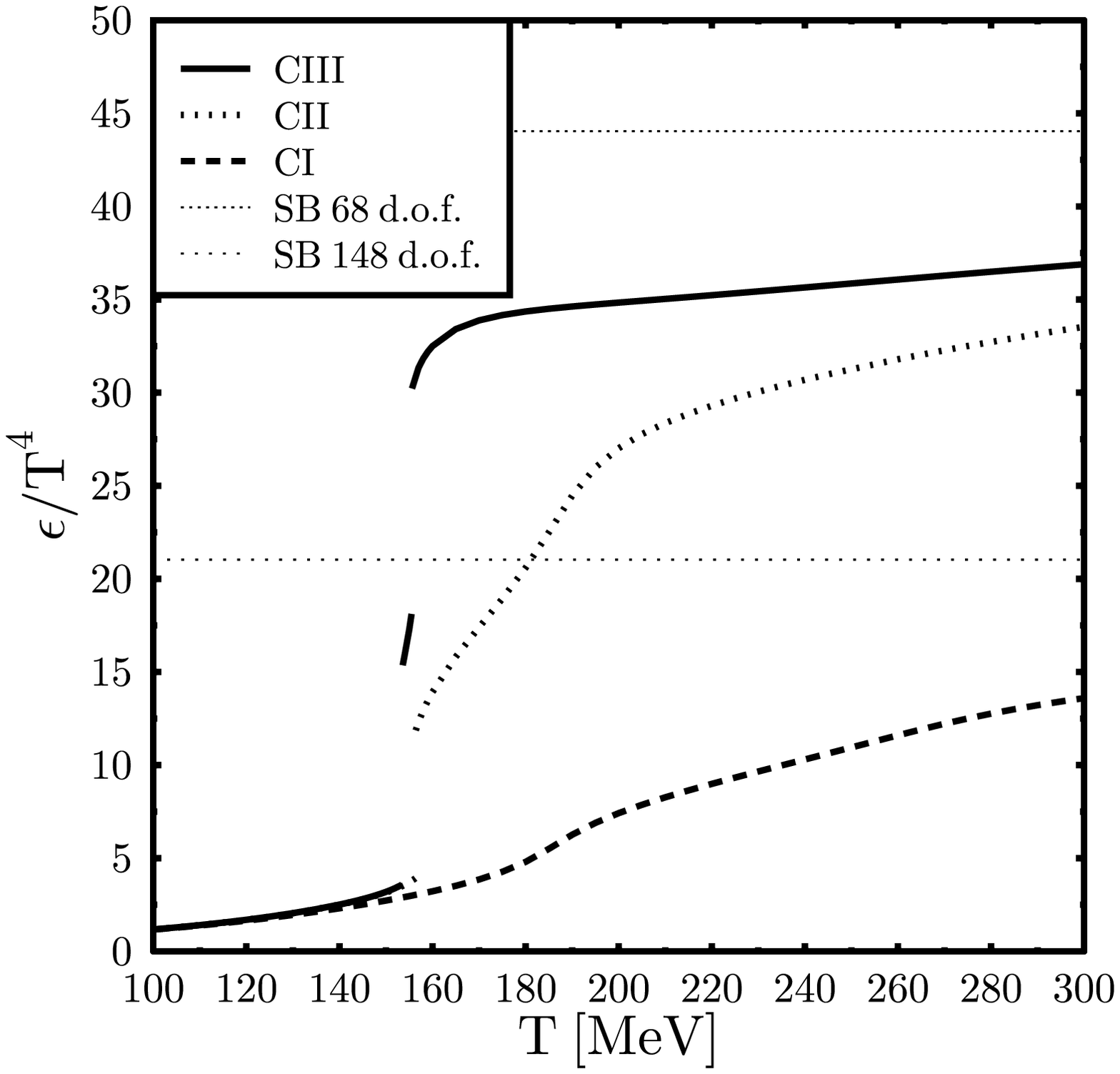}}
\parbox[b]{8cm}{
\includegraphics[width = 9cm]{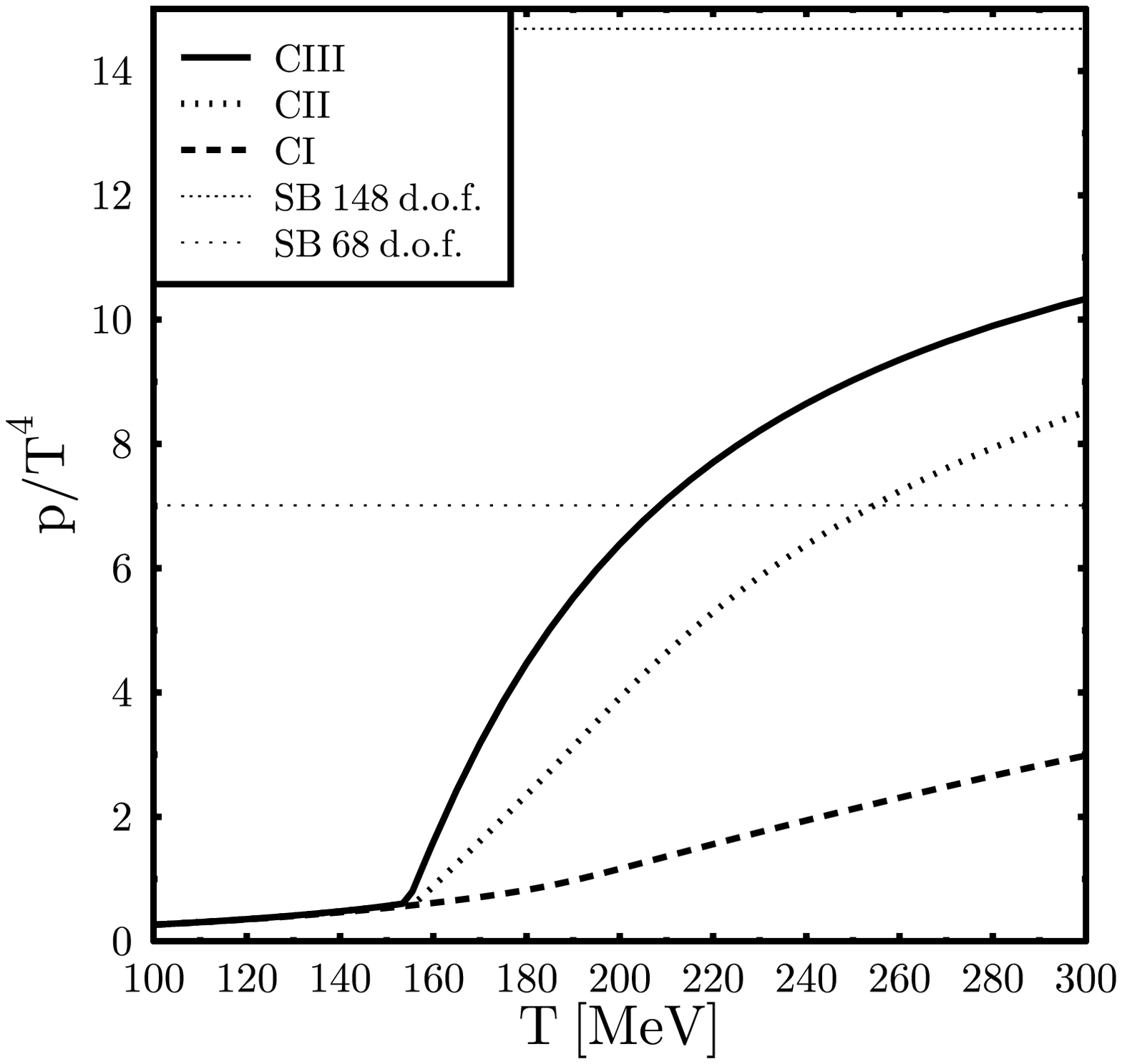}}}
\vspace*{-0.5cm}
\caption{\textrm {$\epsilon/T^4$ and $p/T^4$ for the three different parameter
sets CI, CII, CIII at $\mu_q = \mu_s =0$.
 Depending on the chosen parameters we observe a 
different phase
transition behavior. For CI a smooth crossover occurs. 
In contrast CII leads to a jump in $\epsilon/T^4$ at 
$T \approx 150 \rm{MeV}$ and
a discontinuity in the rise of $P/T^4$ with $T$. Finally, CIII even shows
two discontinuities in $\epsilon/T^4$. The horizontal lines
correspond to the Stefan-Boltzmann limit with and without
the (anti)-baryon decuplet.
\label{et4pvont}}}
\end{figure}

In figure 
\ref{et4pvont} the resulting pressure and energy density are plotted as a 
function of temperature for vanishing chemical potential.
The predicted behavior of the hot hadronic matter
differs significantly for the different parameter sets.
Parameter set CI exhibits a smooth crossover, while   
a first order phase transition is found for parameter set CII. 
Two first order phase transitions are found for parameter set CIII.
This behavior is due to separate jumps in 
the non-strange and the strange condensate.

\begin{figure}[h]
\centerline{\parbox[b]{8cm}{
\includegraphics[width = 9cm]{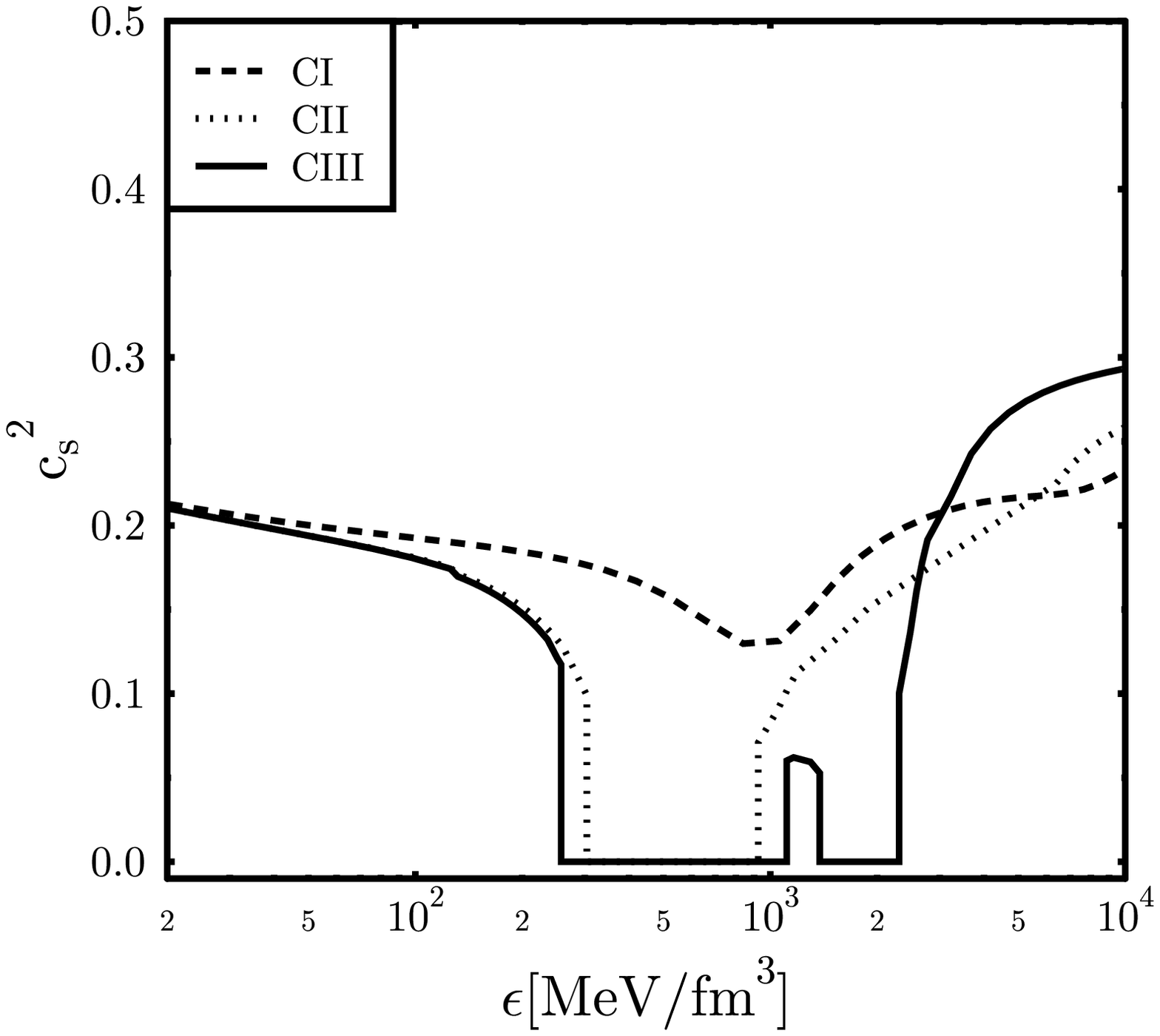}}
\parbox[b]{8cm}{
\includegraphics[width = 9cm]{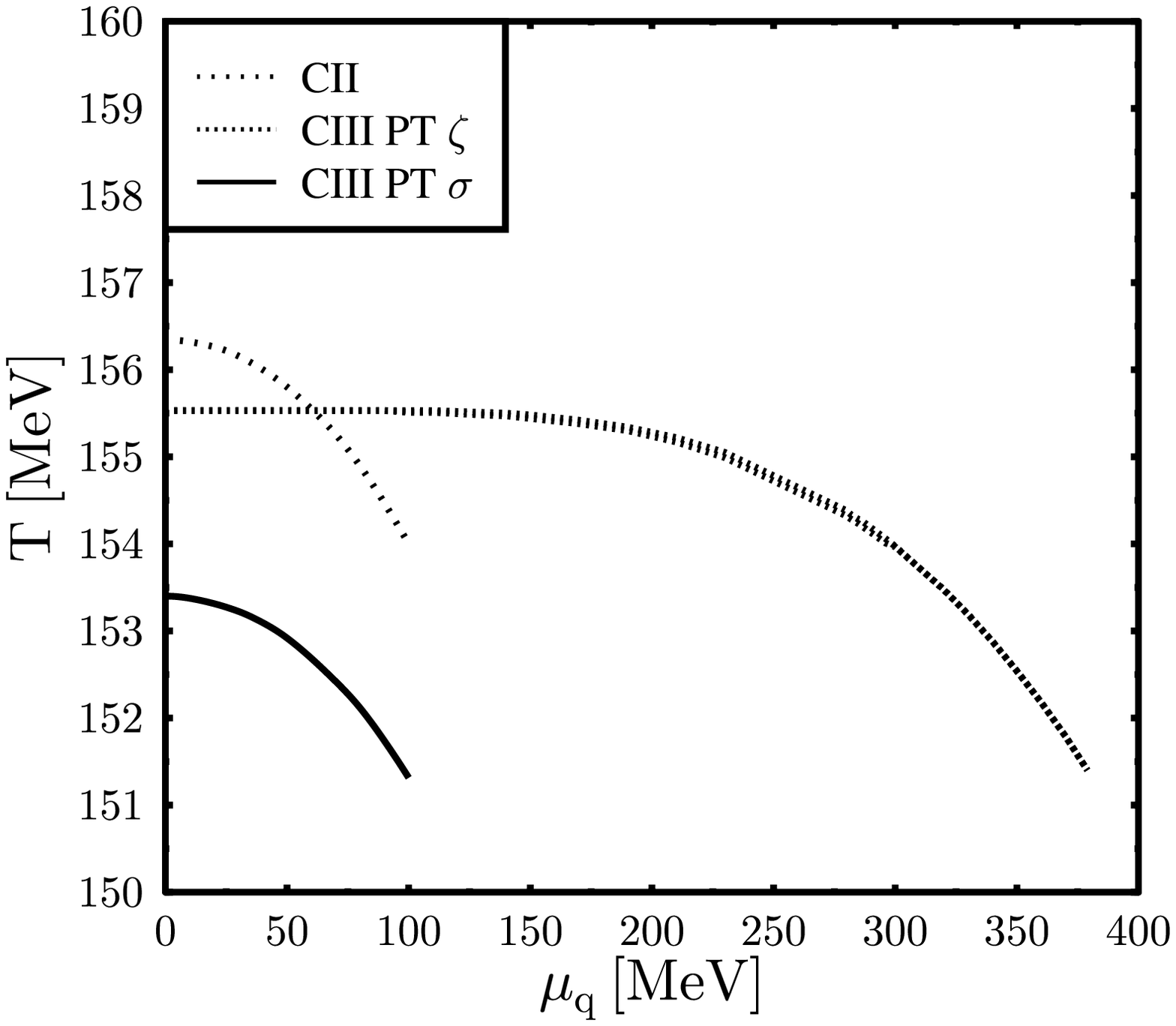}}}
\vspace*{-1cm}
\hfill
\caption{\textrm Left: $c_s^2 \equiv \partial p/\partial \epsilon$ 
for three different equations of state at $\mu_q = \mu_s =0$. 
Right: Phase diagram for the parameter sets CII and CIII for 
$f_s \equiv \rho_s/\rho_B =0$.
The two chemical potentials ($\mu_q,\mu_s$) 
of the system lead to a slight change of the temperature in the 
phase transition region.
 \label{cs2}}
\end{figure}

The resulting velocities of sound are shown in figure \ref{cs2}. 
The crossover EoS shows a decrease of $c_s^2$ around 
$\epsilon = 1 \rm{GeV/fm^3}$. This is due to the strong reduction of the 
baryonic masses around the phase transition region. However, because
the latent heat is zero in the crossover case, $c_s^2$ remains finite. 
In contrast  $c_s^2$ vanishes in the phase transition regions
for CII and CIII (however, it is non-zero if $\mu_q, \mu_s > 0$).
 The latent heat for CII is $\Delta E_{II} \approx 600 \rm{MeV/fm^3}$, while
it is
$\Delta E_{III}\approx 850\rm{MeV/fm^3}+920\rm{MeV/fm^3}=1770\rm{MeV/fm^3}$
for CIII (Both values are for $\mu_q=\mu_s=0$).
Between the two distinct first-order transitions in model III,
$c_s^2$ is non-zero again.
However, this happens in a very narrow interval of energy density, and plays
no significant role in our analysis.
As can be seen from Fig~\ref{cs2} r.h.s., the occurence 
of a first order phase transition depends on the chemical potential. 
For small chemical potential, 
$\mu_q < 100 \rm{MeV}$, 
CIII shows two phase transitions due to the jump in the $\sigma$ and
the $\zeta$ field while CII exhibits one PT due to the jump in the
$\sigma$-field. At higher chemical
potentials, ($100 \rm{MeV} < \mu_q < 370 \rm{MeV}$)  
CIII shows a phase transition due to the 
jump in the $\zeta$ field only. 
Furthermore, since in the $SU(3)$-approach two chemical potentials
($\mu_q,\mu_s$)
have to be considered, the condition $f_s \equiv \rho_s/\rho_B=0$
 does not hold for each
phase in the mixed-phase region, but only for the {\emph total}
strangeness fraction.
This leads to a slight change of the
temperature in the mixed phase. 
For chemical
potential $\mu_q > 370 \rm{MeV}$ there is no phase transition 
for $f_s=0$.
The energy densities and entropy densities in the phase transition
regions are specified in table \ref{enentdensities}.
\begin{table}[h]
\begin{center}
\bt{c|c|c|c|c|c}
     & ${\epsilon^-/\epsilon_0}$ & ${\epsilon^+/\epsilon_0}$  
     & $s^- [fm^{-3}]$ & $s^+ [fm^{-3}]$ 
     & $T_c [\rm{MeV}]$ \\
\hline
CII           & 2.1  & 6.3  & 2.3  & 6.2 & 156.3    \\ 
CIII - 1st PT & 1.7  & 7.6  & 2.0  & 7.5  & 153.4    \\
CIII - 2nd PT & 9.4 & 15.7  & 9.3  & 15.2 & 155.5  \\
 \et
\caption{\label{enentdensities} \textrm{ 
Energy density and entropy density in the phase transition regions
for CII,CIII, $\mu_q = \mu_s = 0$.
The $(-),(+)$ signs stand for values below and above the phase
transition, respectively. $T_c$ denotes the phase transition
temperature.
$\epsilon_0 = 138.45$~MeV/fm$^3$ denotes the energy density of nuclear
matter in the ground state.}}
\end{center}
\end{table}

\begin{figure}[h]
\includegraphics[width = 10cm]{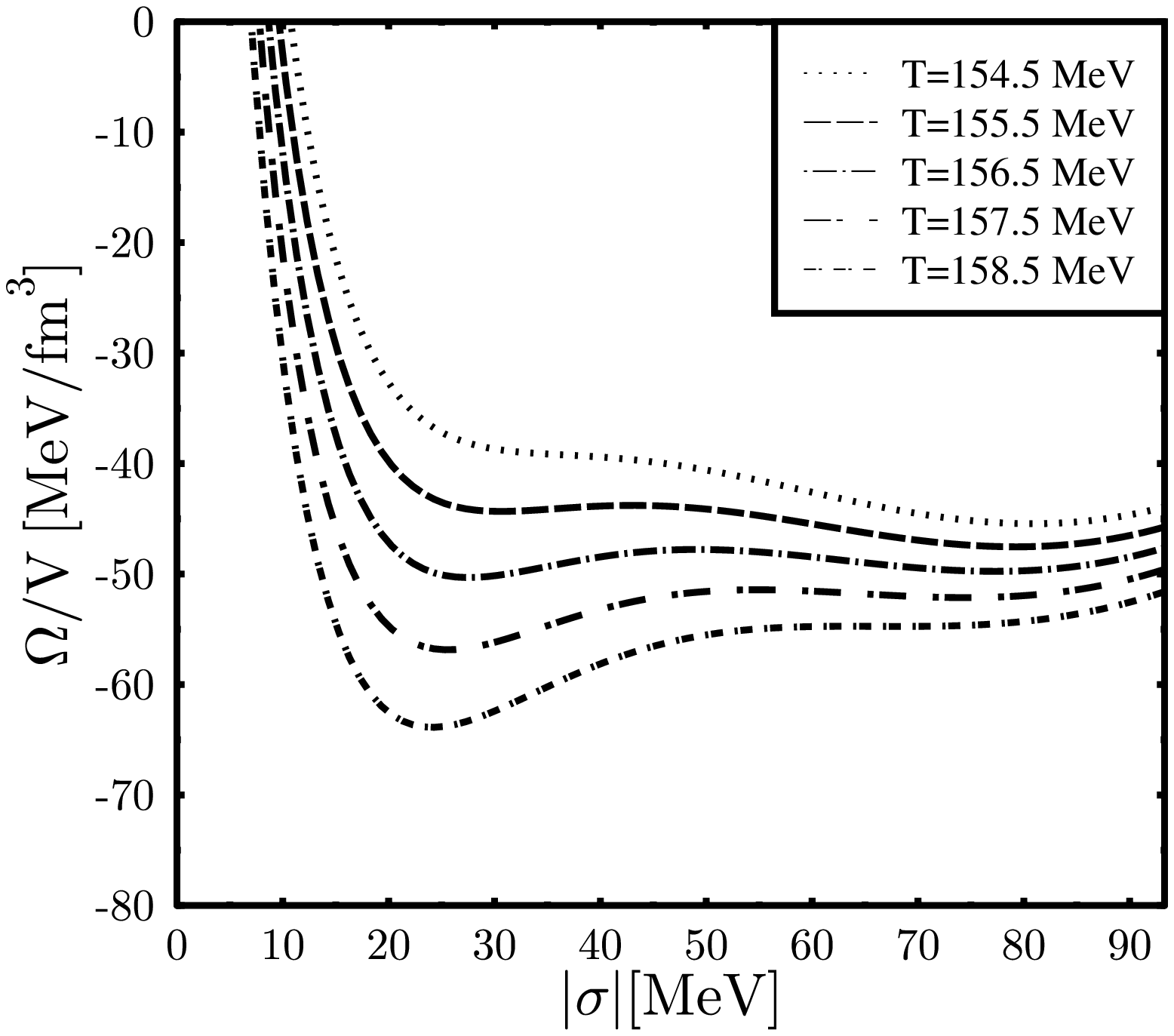}
\caption{\textrm Effective Potential $\Omega/V \equiv -p$ as a function of
the scalar
condensate $\sigma$ around $T_c$. 
For parameter set CII and $\mu_q = \mu_s =0$ 
($\zeta$ has been chosen such as to maximize the pressure 
for given $\sigma$).\label{effpot}}
\end{figure}

The effective thermodynamic potential for parameter set CII around the  
phase transition temperature $T_c$ is depicted in figure \ref{effpot}. 
We observe that
the effective thermodynamic potential varies very rapidly 
around $T_c$. The spinodal
points, i.e. the temperatures at which the inflection points for the
two minima appear, are only $2-3 \%$ off $T_c$. This potential
therefore varies substantially faster than that from the Gross-Neveu model
or from the $SU(2)$ linear sigma model investigated in~\cite{bubble}.
However, the variation of the potential around $T_c$ obtained from our model 
is in the same range as for the model used in~\cite{explosive}, where
the authors showed that such a fast variation of the effective
potential around $T_c$ might lead to explosive behavior via rapid
spinodal decomposition (as opposed to an adiabatic phase transition).
This questions the applicability of
our approach of equilibrium hydrodynamics.
However, as a first approximation, we study the effects dynamically, 
assuming that local equilibrium does hold,
i.e.\ that the mean fields in fact assume the value of
the global minimum of the potential, and that at the critical
temperature two phases (corresponding to the two
minima of the effective potential) coexist.

\subsection{Initial Conditions}

The initial conditions in scaling hydrodynamics are specified on a
proper-time hyperbola $\tau=\tau_i$.
On that space-like hypersurface one has to specify
the entropy per net baryon and the net baryon
rapidity density at midrapidity, $dN_B/dy$.
A model with an MIT bag model equation of state~\cite{mitbag} for the high
temperature phase and an ideal hadron gas in the low-temperature
region can reproduce both \cite{DumRi},  
the measured transverse energy at midrapidity, and the $p_T$-spectra
of a variety of hadrons at $\sqrt{s}=17.4A$~GeV (CERN-SPS energy),
assuming the standard thermalization
(proper) time $\tau_i=1$~fm/c, and a specific entropy
of $s/\rho_B=45$ and a net baryon rapidity density $dN_B/dy=80$.
This value for $s/\rho_B$ is also in good agreement with the
measured relative abundances of hadrons~\cite{Rafelski}.
The initial net baryon density follows as 
$\rho_B = 4.5 \rho_0$. The
corresponding values of $\epsilon_i, T, \mu_q$ and $\mu_s$ 
(q- and s-quark chemical potential respectively)
for the various chiral EoS are listed in
table~\ref{evalues}.
\begin{table}[h] 
\begin{center}
\bt{cc|c|c|c|c|c}
     &  & $\epsilon/\epsilon_0$ & $p/\epsilon_0$ & 
     $T [\rm{MeV}]$ & $\mu_q [\rm{MeV}]$  & 
       $\mu_s [\rm{MeV}]$ \\
\hline
SPS  & CI   & 49.2  & 10.5  & 256.0  & 236.2 & 133.0  \\
     & CII  & 40.2  & 6.6   & 197.0  & 241.3 & 58.6   \\
     & CIII & 37.3  & 5.9   & 180.6  & 246.6 & 36.4   \\
RHIC & CI   & 127.9  & 29.3  & 313.3  & 138.4 & 95.8  \\
     & CII  & 100.4  & 21.4  & 242.0  & 151.6 & 60.9   \\
     & CIII & 93.7   & 22.0  & 230.0  & 154.6 & 53.0   \\
 \et
\caption{\label{evalues} \textrm{ 
Initial conditions for the three chiral EoS,
corresponding to
$s/\rho_B = 45$ and $dN_B/dy=80$ for CERN-SPS energy and
$s/\rho_B = 200$ and $dN_B/dy=25$ for BNL-RHIC energy.
$\epsilon_0 = 138.45$~MeV/fm$^3$ is the energy density of nuclear
matter in the ground state.
Here, $\epsilon$ and $s/\rho_B$ denote the average values at midrapidity at 
the initial time $\tau_i$, i.e.\ the mean of the respective transverse 
distribution. The other quantities have been computed from those average 
values for $\epsilon$ and $s/\rho_B$, using the corresponding EoS.
}}
\end{center}
\end{table}

The initial net baryon density is independent of
the underlying EoS because the continuity equation for the net baryon
current in~(\ref{Hydro}) does not involve the pressure $p$ explicitly. 

Due to the higher density at midrapidity,
thermalization may be faster at BNL-RHIC energies -- following
\cite{RiGy} 
we assume $\tau_i=0.6$~fm/c. Various microscopic models, e.g.
PCM \cite{pcm-ini}, RQMD \cite{rqmd-ini}, FRITIOF 7.02
\cite{frit-ini},
and HIJING/B \cite{hijing-ini}, predict a net baryon rapidity density
of $dN_B/dy \approx 20-35$ and specific entropy 
of $s/\rho_B \approx 150-250$ in central Au+Au at 
$\sqrt{s}= 130 \rm{AGeV}$ at midrapidity. We will employ    
$s/\rho_B=200$ and $dN_B/dy$=25. The resulting baryon density at
midrapidity is $\rho_i = 2.3 \rho_0$.
Hadron multiplicity ratios at midrapidity
can be described with these initial conditions~\cite{RHICratios}. 
The energy density and baryon density are initially distributed in the
transverse plane according to a so-called ``wounded nucleon'' distribution
with transverse radius $R_T=6$~fm. For further details, we refer to
refs.~\cite{DumRi,soff}.
As seen from table~\ref{evalues}, the initial energy density more than doubles
when going from CERN-SPS energy to BNL-RHIC energy. The initial temperature
increases by about 50~MeV, while the initial chemical potential for $u$, $d$
quarks decreases by about 100~MeV, in all cases. Note that for a bag model
EoS the chemical potential for $s$-quarks vanishes because of strangeness
neutrality in the QGP phase, see e.g.\ \cite{DumRi}. 
Strangeness neutrality is a global
constraint, only \cite{distillation}. 
Within a mixed phase, however, the individual phases
may adopt non-zero values for $f_s$.
In a hadronic model,   
the hyperons contain non-strange
quarks and adopt a finite chemical
potential if $\mu_q \ne 0$. 
Therefore, the hyperon vector density is positive 
at finite temperature. This surplus of strange 
quarks contained in the hyperons is balanced by the anti-strange quarks in
strange mesons. This leads to a finite strangeness chemical potential
$\mu_s$, which is adjusted to yield $n_{s}=n_{\bar{s}}$. Here 
$n_{s}, n_{\bar{s}}$ denote the total number of strange and anti-strange 
quarks in the system, respectively.
As already discussed in \cite{distillation}, in the mixed phase only the {\textrm 
total} strangeness fraction $f_s$ vanishes, while each of 
the two coexisting phases does, in general, carry net strangeness.
Furthermore for the case of a strong first order phase transition
the evaporation of pions and kaons and strangeness distillation 
\cite{distillation} should be studied, since these 
influence the unlike particle correlations 
(e.g. $K^+/K^-$, see \cite{soff-ard}).

\section{Results} \label{sec_results}
\subsection{Hypersurfaces} \label{hypersur}
\begin{figure}[hb]
\vspace*{-.5cm}
\includegraphics[width = 16cm]{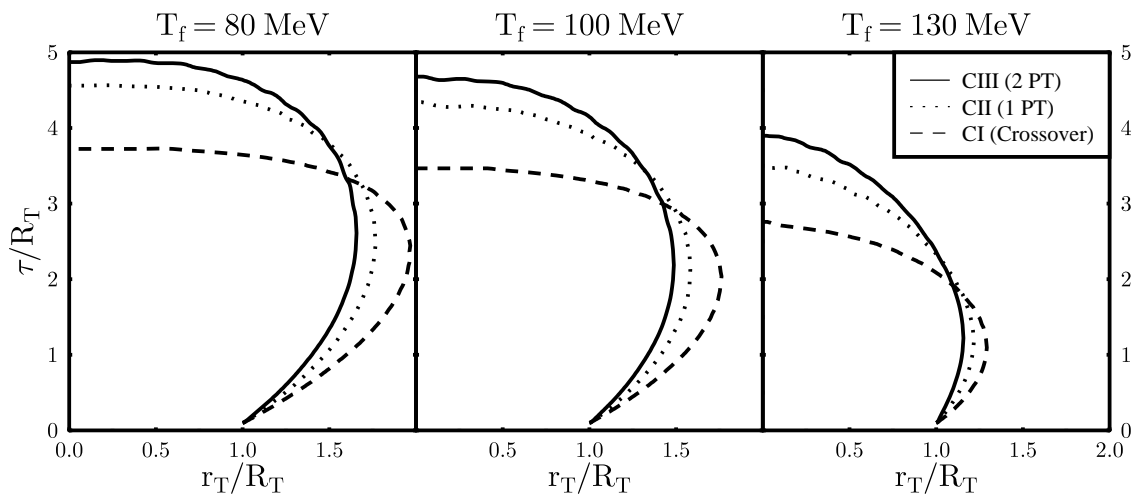}
\caption{\textrm Hypersurfaces $T=T_f$ for the three chiral EoS.
This figure corresponds to initial conditions as appropriate for
central Au+Au collisions at BNL-RHIC energy ($\sqrt{s}=130 \rm{A GeV}$).
\label{spacetime-rhic}}
\end{figure}

Before presenting results on pion correlations, in this section we shall
discuss the effects from varying the latent heat in the EoS on the
space-time evolution of the hadronic fluid. 
Qualitatively, the same
effects are observed for both sets of initial conditions, and we
therefore show only the results corresponding to the BNL-RHIC case.
Figure \ref{spacetime-rhic}
shows the calculated 
hypersurfaces at fixed temperature, $T=T_f$, in the transverse
plane at $\eta=0$ for the three chiral EoS.

Comparing the freeze-out curves for the different 
equations of state one finds that the different phase-transition
behavior of the three parameter sets is reflected in
the space-time evolution of the system. In case I (crossover)
the time until freeze-out is shorter than in case III
(two first order phase transitions). This is not surprising, 
since the occurence of a mixed phase prolongs the expansion time. 
This is due to the above-mentioned drop of the speed of sound, $c_s$.
The fastest expansion is obtained for
a crossover with no latent heat and, accordingly, no discontinuity in the 
entropy density. 

The different space-time evolutions for the three EoS is most obvious for 
$T_{f} = 80 \rm{MeV} $ and $T_{f} = 100 \rm{MeV} $ but can also be seen for 
$T_{f} = 130 \rm{MeV} $. 
For CERN-SPS energies we obtained similar results, with only slightly
smaller liefetimes and radial extensions.

\subsection{Two particle correlations}
To calculate the two particle correlation function we use the method
developed in ~\cite{Shuryak:1973kq,pratt86}. We measure the coincidence
probability $P(\mathbf{p_1},\mathbf{p_2})$ 
of two identical particles with momenta $\mathbf{p_1},\mathbf{p_2}$
relative to the probability of detecting uncorrelated particles.
The inverse width of the
correlation function in out-direction ($R_{out}$) 
is proportional to the
duration of the particle emission, i.e.\ to the lifetime of the
source  \cite{pratt,pratt86}. 
Analogously, the inverse width of the correlation function in
side-direction ($R_{side}$) is a measure for 
the (transverse) size of the source. 
Using a Gaussian fit one can relate the inverse widths of the
correlation functions to radius parameters. 
It was pointed
out in \cite{RiGy} that both for model calculation as well as for 
experimental data it is tedious, if not impossible, to relate 
$R_{side}$ and $R_{out}$ to the real source size and lifetime. 
However, the
ratio $R_{out}/R_{side}$ can be used as a  measure for the
lifetime of the system.

The HBT radii shown below are obtained as follows. We assume that the
pion correlation function is determined on a hypersurface of given temperature
$T_f$, where the pion mean free path
supposedly becomes too large to maintain local equilibrium.
As already mentioned above, at present we refrain from a detailed study
of transport coefficients of our model. Rather, our approach shall be
more pragmatic, and we shall consider $T_f$ as a free parameter.

On the $T=T_f$ hypersurface, the two-particle correlation function is given
by~\cite{schlei,RiGy}
\be \label{C2hadroniz}
C_2({\mathbf p_1},{\mathbf p_2}) = 1 + \frac{1}{{\cal N}}
   \left|
\int{\rm d}\sigma\cdot K e^{i\sigma\cdot q} f\left(u\cdot K/T\right)
   \right|^2
\ee
The normalization factor ${\cal N}$ is given by the product of the
invariant single-particle inclusive distributions of the pions evaluated at
momenta ${\mathbf p_1}$ and ${\mathbf p_2}$, respectively. $u^\mu$
denotes the four-velocity of the fluid on the   $T=T_f$
surface, $\sigma^\mu$; $K^\mu=(p_1^\mu+p_2^\mu)/2$,
$q^\mu=p_1^\mu-p_2^\mu$ are the average four-momentum and the
relative four-momentum of the pion pair, respectively. 
For midrapidity pions, $K_\parallel=q_\parallel=0$.
Thus, for the cylindrical geometry, the correlation
function depends on three variables only;
that is, the $out$ and $side$ components of $q$, and the transverse
momentum of the pion pair, $K_T$.
In~(\ref{C2hadroniz}),     
$f$ denotes the local distribution function of pions in momentum
space, at a temperature $T_f$. For simplicity, we shall assume 
a thermal distribution function and neglect the interaction energy of
the pions, which amounts to only a $\sim5\%$ correction relative to the
vacuum mass of the pion. From $C_2(q_{out},q_{side},K_T)$ we determine
the HBT radii as
$R_{\rm out} = \sqrt{\ln 2}/q_{\rm out}^*$ 
and $R_{\rm side}= \sqrt{\ln 2}/q_{\rm side}^*$,
where $q_{\rm out}^*$, $q_{\rm side}^*$ are defined by 
$C_2(q_{\rm out}^*,q_{\rm side}=0)=
C_2(q_{\rm side}^*,q_{\rm out}=0)=3/2$. 

\begin{figure}[hb]
\vspace*{-0.8cm}
\includegraphics[height = 12cm]{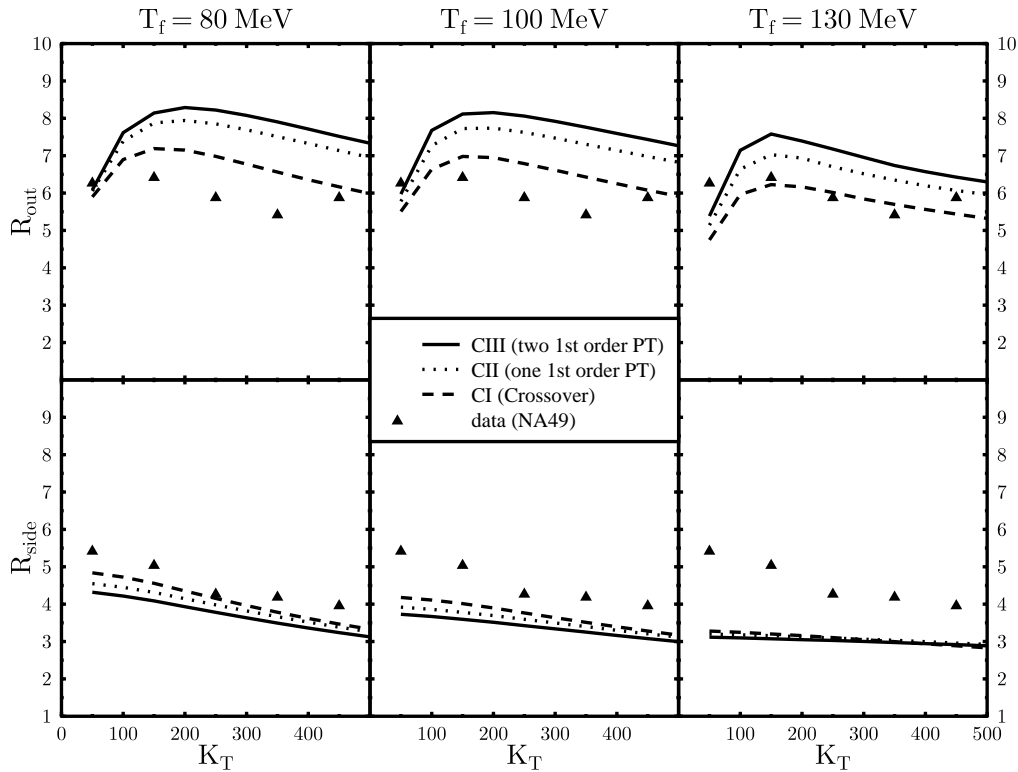}
\vspace*{-1cm}
\caption{\textrm {$R_{side}$ and $R_{out}$ as a function of $K_T$ at SPS.
$T_{f} = 80,100, 130 \rm{MeV}$.
\label{rsro-sps}}}
\end{figure}

In Fig.~\ref{rsro-sps} we show the resulting
HBT-radii $R_{side}$ and $R_{out}$ for central Pb+Pb collisions
at SPS energy ($\sqrt{s}/A=17.4$~GeV),
and compare to recent preliminary
data obtained by the NA49 collaboration~\cite{blume01}.
Of course, in view of the approximations mentioned above such a comparison
should be interpreted with care.
At $T_f=130$~MeV, $R_{out}$ is reproduced reasonably
well.
In particular, it appears that the EoS with the largest latent heat 
overestimates
$R_{out}$. This is rather similar to the bag model EoS~\cite{RiGy,soff}.
Note that $R_{out}$ describes the size of the
source folded with the mean emission
duration~\cite{wiedemannrep,pratt86,schlei,RiGy,soff,pratt}.
The average radius of the pion source
decreases with increasing latent heat, but the emission duration increases.
Integrating over the
emission surface, Fig.~\ref{rsro-sps} shows that for $K_T\ge50$~MeV
the latter dominates in case of longitudinal scaling expansion, and $R_{out}$
increases with the latent heat of the chiral transition.
The EoS with vanishing or small latent heat is 
closest to the data.

At $T_f=80$~MeV and $T_f = 100$~MeV, the pion source
has expanded further 
and hence $R_{out}$ is larger. At large $K_T$, the EoS with first-order
phase transition predicts too large values for $R_{out}$ for both
values of $T_f$. At small transverse
momenta, on the other hand, all three EoS describe the data better than for
the high freeze out temperature. 
This observation is in agreement with the
results of ref.~\cite{soff}, which shows that due to dissipative effects
particles which suffer soft hadronic rescatterings freeze out at much 
later times than particles subject to harder
interactions. We can not account for that effect within our ideal-fluid
model, but it can be mimicked by choosing a lower $T_f$ at
smaller $K_T$. In any case, our main focus is on effects from
the chiral phase transition. Our results suggest that a weakly first-order
transition, or a smooth crossover, can give a better description of
$R_{out}$ than a phase transition with large latent heat (as in the bag model).

$R_{side}$ measures the geometric size of the pion source in the
transverse plane \footnote{More precisely, it corresponds
to the scale of spatial homogeneity~\cite{Makhlin:1988gm}.}, and does not
depend on the emission duration
~\cite{wiedemannrep,pratt86,schlei,RiGy,soff,pratt}.
First, we note that the effective source radius depends only very weakly
on the latent heat for the transition, in particular for large $T_f$.
This is in accord with the space-time evolution as described in
\ref{hypersur}. At small $K_T$, $R_{side}$ decreases slightly with
the latent heat. However, 
for all three EoS, $R_{side}$ comes out too small.
Only for $T_f=80$~MeV an reasonable description is obtained. 
This could be partly due to the neglect of resonance decays, which
form a ``halo'' surrounding the direct pion
source~\cite{gyulassy89,schlei,wiedemannrep,Zimany}, and increase 
its effective size.
On the other hand, the resonance decays would also tend to increase
$R_{out}$. As discussed in~\cite{RiGy}, a reasonable measure for the
emission duration of pions therefore is the ratio $R_{out}/R_{side}$,
which is also less affected by resonance decays.
\begin{figure}[hb]
\vspace*{-0.8cm}
\includegraphics[width = 16cm, height= 12cm]{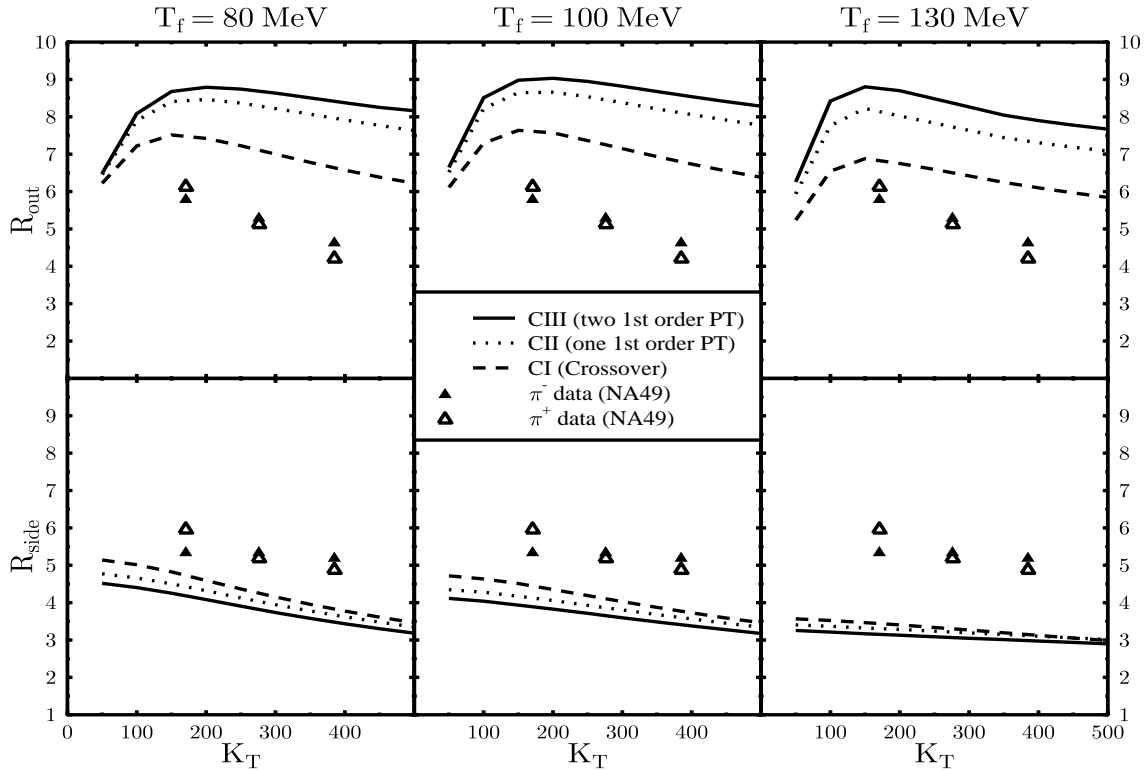}
\vspace*{-1cm}
\caption{\textrm {$R_{side}$ and $R_{out}$ as a function of $K_T$ at RHIC.
$T_{f} = 80, 100, 130 \rm{MeV}$.
\label{rsro-rhic}}}
\end{figure}
In Fig.~\ref{rsro-rhic} we show the results for initial conditions
appropriate for BNL-RHIC energies. Both radii increase as compared to the
lower SPS energy. That is because the initial entropy density is significantly
larger. Thus the system takes longer to cool down to $T_f$, 
and the system has
more time to expand in the transverse direction. 
For example, at $K_T=500$~MeV, $R_{out}$ increases by about 1~fm
for the EoS with a strongly first-order phase transition.
$R_{side}$ increases even less. This is in contrast to an EoS with only
pions in the hadronic phase~\cite{pratt,pratt86,RiGy}, where the ratio
of entropies
of the two thermodynamic phases is very large at $T_c$.
The very moderate increase of the radii from SPS to RHIC energy is in
agreement with the results from STAR for Au+Au collisions at
RHIC~\cite{STAR_HBT}.
On the other hand, as already discussed above, the ``geometric
size'' of the source, $R_{side}$, is too small. As at SPS energy, this could
be due to decays of resonances. We shall therefore discuss next the
behavior of the ratio $R_{out}/R_{side}$ with $K_T$, which is less
affected by decays~\cite{RiGy}, and which is a good measure for the
lifetime of the pion source.

\begin{figure}[hb]
\vspace*{-.8cm}
\includegraphics[width = 16cm]{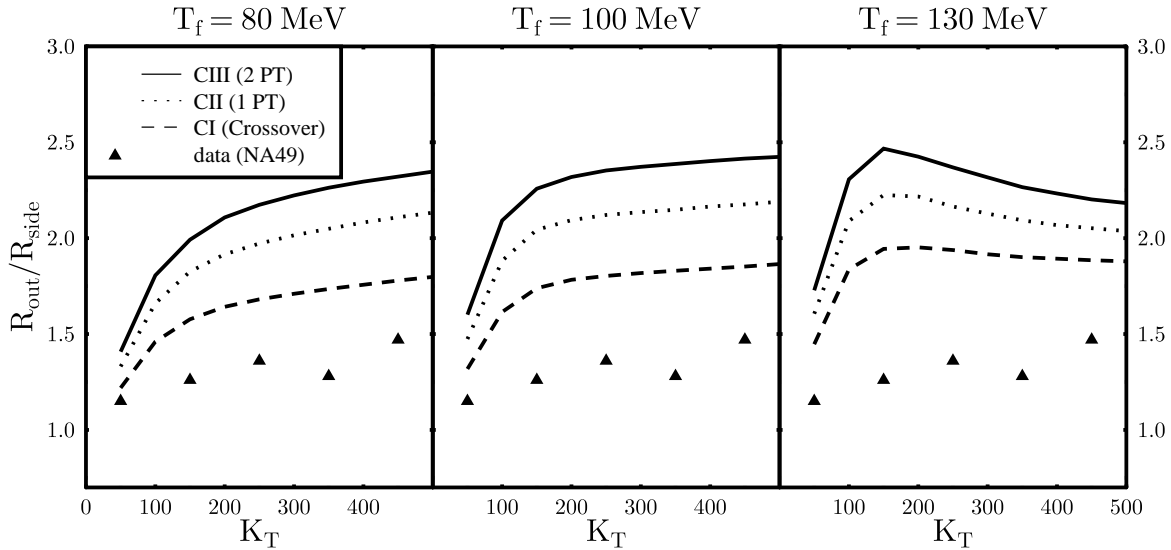}
\vspace*{-1cm}
\caption{\textrm {$R_{out}/R_{side}$ as a function of $K_T$ at SPS. \label{hbt-sps}}}
\end{figure}

\begin{figure}[h]
\vspace*{-.8cm}
\includegraphics[width = 16cm]{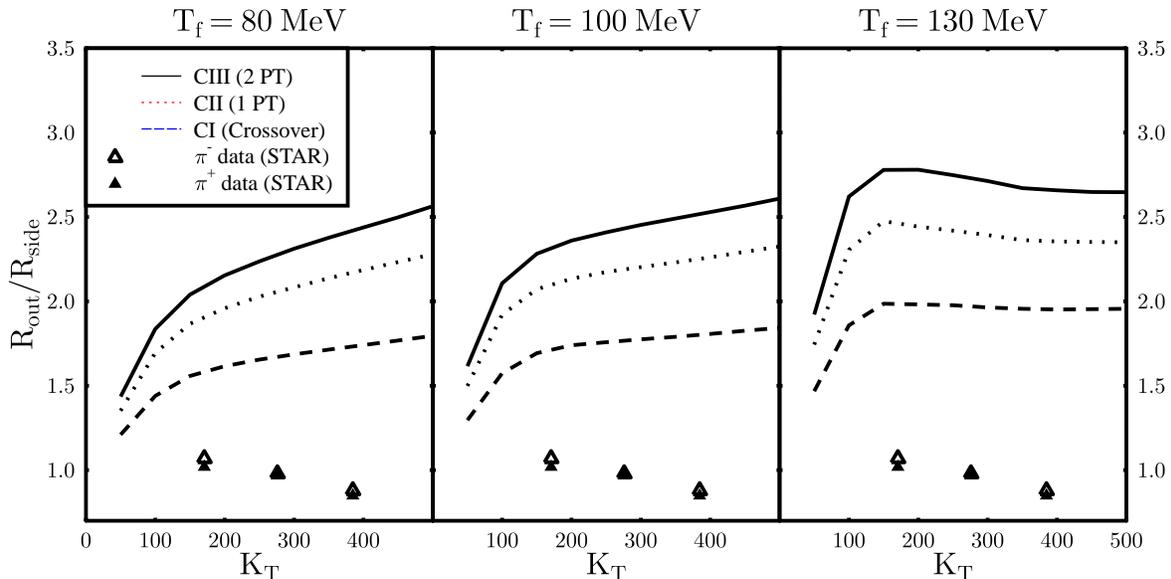}
\vspace*{-1.2cm}
\caption{\textrm {$R_{out}/R_{side}$ as a function of $K_T$ at RHIC.\label{hbt-rhic}}}
\end{figure}

Figures \ref{hbt-sps} and
\ref{hbt-rhic} show the 
experimentally measured ratio $R_{out}/R_{side}$ as a function 
of $K_T$ for the three different equations of state for SPS and RHIC
energies.  
One observes that at both energies the shortest lifetime of the system
emerges from the EoS featuring a crossover (CI), 
while the slowest expansion results
from the EoS with largest latent heat (CIII). 
At SPS energy, the data~\cite{blume01} yield a slowly rising
$R_{out}/R_{side}$ ratio. This is obtained for all three EoS, if the
freeze-out temperature is low, $T_f=80,100$~MeV. Of course, the absolute
value of $R_{out}$ is too large, while $R_{side}$ is too small, and so
the ratio
comes out way too large. As is obvious from the figures, we are not able to
reproduce the data for $R_{out}/R_{side}$, though a small or even vanishing
latent heat and a smaller freeze-out temperature improve the
picture. 
Turning to RHIC, we see that the predicted general behavior of
$R_{out}/R_{side}$ is similar as at the SPS, 
except for a slight overall increase of that ratio.
That is because the larger initial entropy
per baryon, which is deduced from the larger $\pi/p$ and $\bar{p}/p$
ratios, increases the lifetime of the system slightly.
On the other hand, the STAR data \cite{STAR_HBT} show 
$R_{out}/R_{side}\simeq1$, and a 
decrease with $K_T$. This behavior can evidently not be reproduced for
low $T_f$. Larger freeze-out temperatures, 
$T_f=130$~MeV, lead to flat, or even slightly decreasing
$R_{out}/R_{side}$. Nevertheless, $R_{out}/R_{side}$
is about a factor of 2 higher for $K_T\ge100$~MeV than seen in the
STAR data.

\section{Conclusion} \label{sec_summary}
The space-time evolution of
ultra-realtivistic heavy ion collisions at SPS and RHIC energies
has been studied 
within hydrodynamic simulations using various EoS obtained from a chiral
$SU(3) \times SU(3)$ model. 
HBT radii have been calculated and compared to data from the NA49
collaboration and the STAR collaboration. The influence of 
different orders of the chiral phase
transition and the underlying EoS have been discussed.

A small latent heat, i.e.\ a weak
first-order chiral phase transition or even a smooth crossover, leads
to larger emission regions and smaller emission duration, as
well as to larger $R_{side}$ and smaller $R_{out}$ 
HBT radii than a strong first
order transition as for example assumed in the bag model.

In almost all
cases, we observe that the results obtained with a crossover EoS 
are closest to the experimental data.
However, a quantitative description of the data, both at SPS energy
as well as at RHIC energy, 
is not possible within our present ideal-fluid approach with
longitudinal scaling flow, employing the various SU(3) chiral EoS.

Apparently, conclusions can only be drawn after 
considerable improvements on various
aspects of the description of high-energy heavy-ion collisions.

In particular, the effective chiral-SU(3) potential
is rather rapidly varying around $T_c$. Therefore, a non-equilibrium
description, accounting for supercooling effects and/or
rapid spinodal decomposition, might be in order~\cite{bubble,explosive}.
In fact, it has been argued that the decay of a
droplet of chirally symmetric matter from a region of negative pressure
may yield smaller values for the HBT radii, as well as a smaller ratio
$R_{out}/R_{side}$~\cite{cc1}\footnote{Note, however, that within our
effective theory for chiral symmetry restoration the spinodal instability
occurs before negative pressure is reached, see Fig.~\ref{effpot}.}.
Such considerations are out of the scope of the present manuscript
but will be studied in detail in the future.
More realistic freeze-out descriptions 
\cite{soff,bugaev} may improve the results.
However, as dissipative effects are expected to prolong the lifetime of the
pion source even more, it appears very likely that a quasi-adiabatic
first-order phase transition with large latent heat, for which
a hydrodynamic description should be adequate, can not describe
the pion HBT data from CERN-SPS and from BNL-RHIC. This observation
may be viewed as an experimental confirmation of the predictions 
from lattice QCD~\cite{lattice_chi,Laerm}, which do not show a large 
latent heat. 

The nature of the chiral symmetry restoration
will be better understood by analyzing forthcoming experimental 
data from RHIC.
For example, correlations among kaons, protons, and non-identical particles
can be analysed~\cite{Panitkin_QM01}. Excitation functions between
CERN-SPS energy ($\sqrt{s}=17.4A$~GeV)
and the present BNL-RHIC energy ($\sqrt{s}=130-200$~AGeV) would be 
extremely useful to  
provide a more detailed view
of the behavior of the correlation functions.
The excitation functions of source sizes 
and lifetimes and 
also  so-called ``azimuthally sensitive'' HBT-analysis~\cite{az_HBT} 
could be
useful to obtain a complete picture of the phase transition 
via the structure of the
pion source in space-time. With these data emerging, we hope that it
will be possible to obtain a deeper understanding of the 
QCD phase transition in high-energy heavy-ion collisions.
\section*{Acknowledgement}
We would like to thank A.~Dumitru and D.~Rischke 
for numerous discussions and
for providing the 
relativistic hydrodynamics and the appropriate correlation
function codes. We also thank K.~Bugaev, I. Mishustin and L. Satarov
for discussions.
This work was supported by Deutsche 
Forschungsgemeinschaft (DFG), Gesellschaft f\"ur Schwerionenforschung
(GSI), Bundesministerium f\"ur Bildung und Forschung (BMBF),
Graduiertenkolleg Theoretische und experimentelle Schwerionenphysik
and by the U.S. Department of Energy, Nuclear Physics
Division (Contract No. W-31-109-Eng-38).

\bibliographystyle{prsty}

\end{document}

%% file: define.tex
\newcommand{\bea}{\begin{eqnarray}}
\newcommand{\eea}{\end{eqnarray}}
\newcommand{\be}{\begin{equation}}
\newcommand{\ee}{\end{equation}}
\newcommand{\bt}{\begin{tabular}}
\newcommand{\et}{\end{tabular}}

\newcommand{\no}{\nonumber}

\newcommand{\beas}{\begin{eqnarray*}}
\newcommand{\eeas}{\end{eqnarray*}}
\def\Journal#1#2#3#4{{#1} {\bf #2}, #3 (#4)}

\def\NP{{Nucl. Phys.}}
\def\PL{{Phys. Lett.} B}

\def\PRD{{Phys. Rev.} D}

\def\MPLA{{Mod. Phys. Lett. A}}